\documentclass[aps,twocolumn,superscriptaddress,amsmath,amssymb,nofootinbib]{revtex4-1}
\usepackage{graphicx}
\usepackage{multirow,dcolumn}
\usepackage{bm}
\usepackage{txfonts}
\usepackage{hyperref}
\usepackage{soul}

\usepackage{epsf,amsmath,amssymb,verbatim,color,multirow,pifont,graphicx,cleveref,tabularx,cancel,soul}
\usepackage[dvipsnames]{xcolor}
\usepackage{rotating}
\usepackage{amsmath,amsfonts,amssymb,bm,cancel}

\begin{document}
\title{Nonequilibrium phase transition in an open quantum spin system
with long-range interaction}

\author{Minjae Jo}
\affiliation{CCSS, CTP and Department of Physics and Astronomy, 
Seoul National University, Seoul 08826, Korea}

\author{Jaegon Um}
\affiliation{CCSS, CTP and Department of Physics and Astronomy, 
Seoul National University, Seoul 08826, Korea}
\affiliation{BK21PLUS Physics Division, Pohang University of Science and Technology, Pohang 37673, Korea}

\author{B. Kahng}
\email{bkahng@snu.ac.kr}
\affiliation{CCSS, CTP and Department of Physics and Astronomy, 
Seoul National University, Seoul 08826, Korea}


\begin{abstract}
We investigate a nonequilibrium phase transition in a dissipative and coherent quantum spin system using the quantum Langevin equation and mean-field theory. Recently, the quantum contact process (QCP) was theoretically investigated using the Rydberg antiblockade effect, in particular, when the Rydberg atoms were excited in $s$-states so that their interactions were regarded as being between the nearest neighbors. However, when the atoms are excited to $d$-states, the dipole--dipole interactions become effective, and long-range interactions must be considered. Here, we consider a quantum spin model with a long-range QCP, where the branching and coagulation processes are allowed not only for the nearest-neighbor pairs, but also for long-distance pairs, coherently and incoherently. Using the semiclassical approach, we show that the mean-field phase diagram of our long-range model is similar to that of the nearest-neighbor QCP, where the continuous (discontinuous) transition is found in the weak (strong) quantum regime. However, at the tricritical point, we find a new universality class, which was neither that of the QCP at the tricritical point nor that of the classical directed percolation model with long-range interactions. Implementation of the long-range QCP using interacting cold gases is discussed.
\end{abstract}



\maketitle


\section{Introduction}
\label{sec:introduction}
Nonequilibrium phase transitions into an absorbing state have been extensively studied ~\cite{marro, harris, kinzel, ziff, dickman, grassberger, obukhov, cardy,hinrichsen, henkel, odor}. However, in recent years, they have attracted a significant amount of attention because some of these transitions have been experimentally realized in turbulence~\cite{sano1,sano2} and dissipative Rydberg atom quantum systems~\cite{gutierrez}. One of the most robust classes of absorbing transitions is the directed percolation (DP) class~\cite{grassberger,obukhov,cardy,hinrichsen,henkel}, in which the dynamics spreads by a contact process (CP). An active particle becomes inactive at a rate $\gamma$, whereas an inactive particle becomes active at a rate $\kappa$ when it contacts a neighboring active particle. If $\kappa/\gamma$ is small, the system falls into an absorbing state. Otherwise, it is in an active state. The CP model can be used for modeling the epidemic spread of infectious disease and the reaction--diffusion process of interacting particles. On the other hand, the Reggeon field theory reveals the universal properties of the DP class~\cite{grassberger_conjecture, janssen_conjecture}.

The CP can be generalized in various ways. Here, we introduce two cases associated with the main topic of this paper. One is the long-range CP. This process was inspired by disease contagion by long-distance insect flight. We recall a simple lattice model associated with the long-range CP~\cite{LDP1, LDP2, LDP3, LDP4, LDP5, LDP6}, in which the activation process is modified as follows. At a rate $\kappa P({\bf x})$, each active particle activates an inactive particle at distance $|{\bf x}|$ in a random direction. $P({\bf x})$ is thought to follow the power law $P({\bf x}) \sim 1/|{\mathbf x}|^{d+p}$, where $d$ is the spatial dimension, and $p > 0$ is a control parameter. Owing to the long-range interaction, the transition property of the DP class can be changed when $p < p_c$, where $p_c$ depends on the dimension $d$~\cite{LDP2}. When $p > p_c$, the long-range interaction is irrelevant. 
The other variant is the so-called tricritical CP~\cite{ohtsuki1,ohtsuki2,grassberger3,lubeck, windus1, windus2}. 
In this modification, in addition to the ordinary CP, 
an inactive particle becomes active at a rate $\omega$ 
when it contacts two consecutive active particles. This tricritical CP exhibits 
a first-order transition for $\kappa < \omega$ and 
a second-order transition for $\kappa > \omega$. 
Thus, a tricritical point occurs at $\kappa=\omega$ with the 
tricritical directed percolation (TDP) class.

Although the DP class is theoretically well established, experimental realization of DP behavior has been elusive~\cite{hinrichsen_experiment}. It was only recently that two experiments associated with this DP class were implemented~\cite{sano1,sano2,gutierrez}. 
We are particularly interested in the experiment in dissipative quantum systems of Rydberg atoms. An essential factor for realizing the DP class in Rydberg atoms is the antiblockade effect. An inactive spin is activated by detuning the excitation energy so that it is comparable to the energy of interaction with the active spin of the nearest neighbor~\cite{antiblockade1,antiblockade2,antiblockade3}. This is reminiscent of the branching process in the CP. We remark that the antiblockade dynamics can be implemented
incoherently when strong dephasing noise is applied. Then, quantum coherence becomes negligible, and the dynamics is reduced to the classical DP process. When quantum coherence is effective, this case is called the quantum contact process (QCP), and coherent and incoherent CPs can be realized simultaneously~\cite{marcuzzi1,marcuzzi2,buchhold}. Competition between the two types of process leads to the TDP class at the tricritical point. This resembles the behavior of spin glass systems, in which competing interactions between spins generate a negative cubic term of the Landau free energy and a tricritical point. In the strong quantum regime, the system undergoes a discontinuous transition. 

When Rydberg atoms interact via the dipole--dipole interaction, it is natural to consider cold atomic systems with long-range interaction. Similar studies of dipole--dipole interactions were performed in quantum systems associated with several phenomena, for instance, quantum magnetism~\cite{quantumMLR,quantumMLR2,quantumMLR3,quantumMLR4}, Anderson localization~\cite{andersonLR1,andersonLR2,andersonLR3,andersonLR4}, Rydberg energy transport~\cite{rydbergLR1}, and Rydberg blockade~\cite{rydbergBL}. 
However, the long-range Rydberg atom system under the antiblockade condition has not been investigated yet, 
even though the results are expected to contribute to theoretical development of the QCP. 
In this paper, we consider the long-range QCP in the open quantum spin system. 
We set up the Lindblad equation for the density matrix in terms of the Hamiltonian 
with long-range interaction and the dissipators for decay and long-range branching and coagulation. 
Using mean-field (MF) theory, we obtain a phase diagram including absorbing and active states, 
and discontinuous and continuous transition curves with a tricritical point. 
This diagram is similar to that of the classical TDP model. 
However, the continuous transition changes from the ordinary DP to the long-range DP class~\cite{LDP1,LDP5}. 
The TDP transition at the tricritical point also changes. 
We expect it to be in a long-range TDP class corresponding to the TDP; 
however, it has not been explored yet. Using the scaling argument, 
we determine the critical exponents of the long-range TDP in the MF limit. Moreover, we determine the upper critical dimension.

The remainder of this paper is organized as follows. In Sec.~\ref{sec:2}, we derive the quantum Langevin equation of the long-range QCP. The MF equation and its phase diagram are presented in Sec.~\ref{sec:3}, and the scaling behavior and upper critical dimension are presented in Sec.~\ref{sec:4}. Finally, we conclude our work and discuss the relationship between our model and the behavior of interacting cold gases in Sec.~\ref{sec:conclusion}.

\section{Equations of motion for the long-range quantum contact process}

\subsection{Lindblad equation}
\label{sec:Lind}
The Lindblad equation describes a quantum system coupled to the environment
in the context of the Born--Markov approximation~\cite{breuer}. 
We consider a quantum spin model on a $d$-dimensional lattice, where 
each spin state denotes the state of a single atom at a site with 
$\left| \uparrow \rangle\right.$, that is, 
an active state, and $\left| \downarrow \rangle\right.$, that is, an inactive state. 
Interactions between atoms and between atoms and the baths may result in the dynamics of the QCP,
which are described by the Lindblad equation.
The equation is generally composed of 
the Hamiltonian and dissipative terms. Our equation also contains 
the coherent terms for branching and coagulation and incoherent ones for not only 
branching and coagulation, but also decay of active states (see Fig.~1),
and is given by   
\begin{eqnarray}
\label{eq:lindqcp}
\partial_t\hat{\rho}&=&-i\left[ \hat{H}_S,\hat{\rho} \right]
+ \sum_l\left[ \hat{L}^{(d)}_{l}\hat{\rho} \hat{L}^{(d)\dagger}_{l}
-\frac{1}{2} \left\{ \hat{L}^{(d)\dagger}_{l}\hat{L}^{(d)}_{l},\hat{\rho} \right\} \right] \nonumber\\
&&+ \sum_{i=b,c}\sum_{l,m}\left[ \hat{L}^{(i)}_{ml}\,\hat{\rho} \hat{L}^{(i)\dagger}_{ml}
-\frac{1}{2} \left\{ \hat{L}^{(i)\dagger}_{ml}\hat{L}^{(i)}_{ml},\hat{\rho} \right\} \right]\,,
\end{eqnarray}
where the Hamiltonian $\hat{H}_S$ is defined as
\begin{equation}
\hat{H}_S= \omega \sum_{l,m} P(|{\bf x}_m - {\bf x}_l|)
\left( \hat{n}_m \hat{\sigma}^+_l + \hat{n}_m \hat{\sigma}^-_l \right) \,,
\label{eq:HS}
\end{equation}
and the Lindblad jump operators of decay, branching, and coagulation are given by
\begin{eqnarray}
\label{eq:ld}
\hat{L}_{l}^{(d)} &=& \sqrt{\gamma} \hat{\sigma}_{l}^{-} \,, \\
\label{eq:lb}
\hat{L}_{ml}^{(b)}&=&\left[ \kappa P(|{\bf x}_m - {\bf x}_l|)\right]^{1/2} \hat{n}_m\hat{\sigma}^+_{l} \,, \\
\label{eq:lc}
\hat{L}_{ml}^{(c)}&=&\left[ \kappa P(|{\bf x}_m - {\bf x}_l|)\right]^{1/2} \hat{n}_m\hat{\sigma}^-_{l} \,,
\end{eqnarray}
respectively.
Here, $\hat{\sigma}^{+}_l$ and $\hat{\sigma}^{+}_l$ are the raising 
and lowering operators of the spin at site $l$, respectively, 
which are defined in terms of the spin basis as 
$\hat{\sigma}^{+} = \left| \uparrow \rangle \langle \downarrow \right|$ and
$\hat{\sigma}^{-} = \left| \downarrow \rangle \langle \uparrow \right|$. 

\begin{figure}[!t]
\includegraphics[width=0.9\columnwidth]{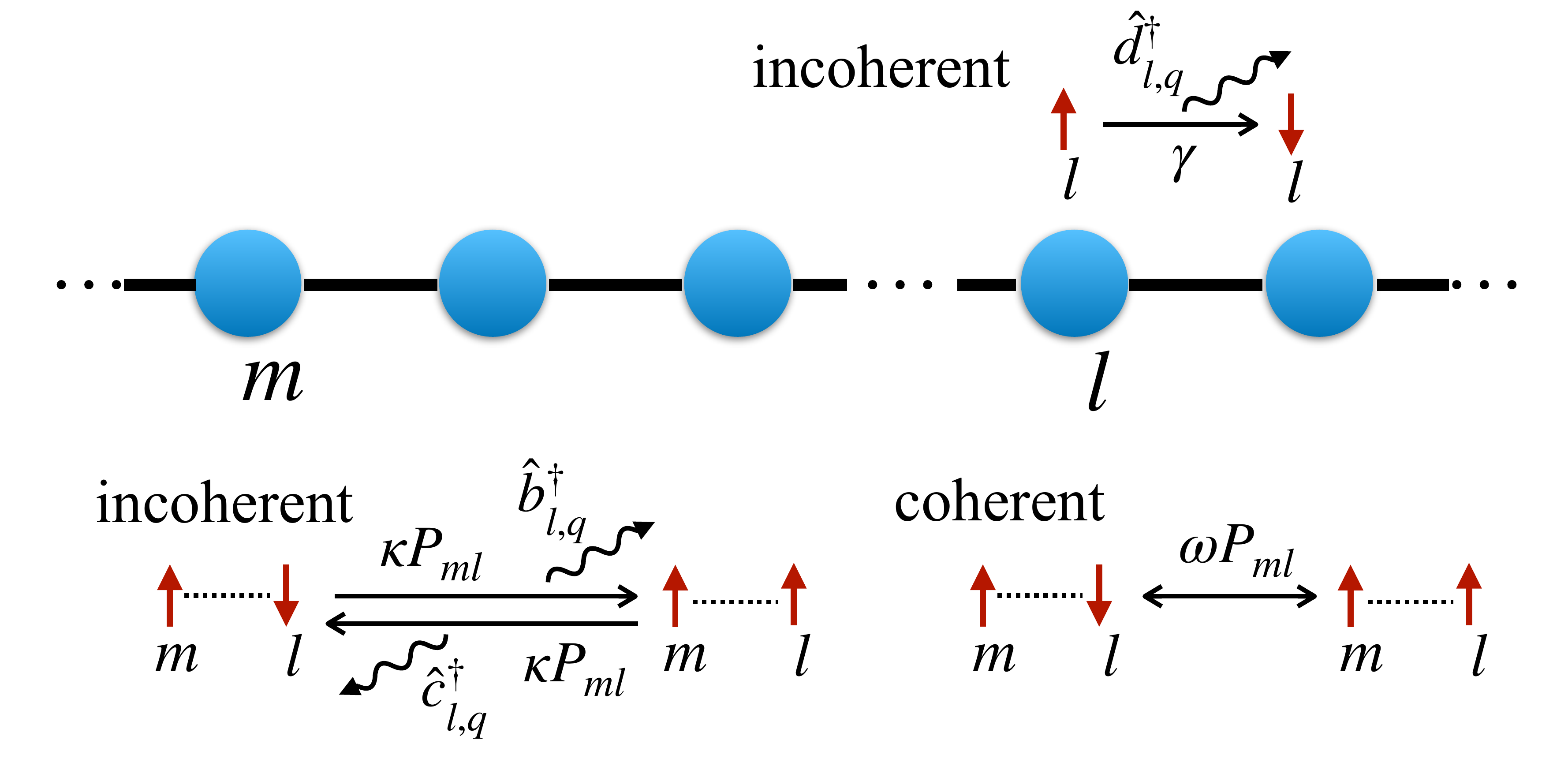}
\caption{Schematic of QCP with long-range interaction in one dimension. 
In this model, there are two incoherent processes and one coherent process, which are represented by the total Hamiltonian in Eq.~\eqref{eq:H}. The incoherent processes are induced by interaction with harmonic baths. The first incoherent process decays each site (denoted as $l$) by raising the harmonic bath's state at the rate $\gamma$, given by the second summation term in Eq.~\eqref{eq:Hd}. The second incoherent process consists of branching and coagulation. Specifically, if site $m$ is in an active state, site $l$ branches (coagulates) at a rate $\kappa P_{ml}$ via directional links (denoted as $ml$) by raising the bath's state, which is given by the second summation term in Eqs.~\eqref{eq:Hb} and~\eqref{eq:Hc}. The rate decreases algebraically as the distance increases. Similarly, the coherent process, which consists of quantum long-range branching and coagulation and involves the off-diagonal elements of the density matrix during the dynamics, is induced by the system Hamiltonian [Eq.~\eqref{eq:HS}].
\label{fig1}}
\end{figure}

Because $\hat{n}_l$ is the number operator of the active state, 
$\hat{n} = \left| \uparrow \rangle \langle \uparrow \right|$,
the composite operator $\hat{n}_m\hat{\sigma}^+_{l}$ or $\hat{n}_m\hat{\sigma}^-_{l}$ 
with $l \neq m$ means that
the active state at site $m$ activates or deactivates the state at $l$, representing the
branching and coagulation processes, as seen in
Eqs.~\eqref{eq:HS}, ~\eqref{eq:lb}, and ~\eqref{eq:lc}. 
Instead, $\hat{L}_{l}^{(d)}$ denotes the decay dynamics of the active state at $l$. 
Therefore, if there is no active state, no further dynamics occurs, implying an absorbing state. 
Note that
$m$ and $l$ need not be a nearest-neighbor pair in the interaction. 
Indeed, $P(|{\bf x}_m - {\bf x}_l|)$ in the dynamic equation [Eq.~\eqref{eq:lindqcp}] represents
the L{\'e}vy distribution, which decays as 
\begin{equation}
\label{eq:levy}
P(|{\bf x}_m - {\bf x}_l|) \sim 1/|{\bf x}_m - {\bf x}_l|^{p+d} \,, 
\end{equation}
and which determines the amplitude of the long-range interaction. 
Here, we set $P(|{\bf x}_m - {\bf x}_l|) =0$ when $m=l$. In addition, the distribution satisfies
the normalization condition,
\begin{equation}
\sum_{m} P(|{\bf x}_m - {\bf x}_l|)
= \sum_{l} P(|{\bf x}_m - {\bf x}_l|) =1 \,. \nonumber 
\end{equation}

It is obvious that the dynamics of populations (the diagonal elements of $\hat{\rho}$) in Eq.~\eqref{eq:lindqcp} in the absence of coherent dynamics is equivalent to the ordinary long-range contact process. Consequently, depending on parameters such as $\kappa$ and $\gamma$, the steady state for $\omega=0$ shows the active or inactive phase, and the transition between them belongs to the long-range DP universality class. If $\omega$ is increased, the coherence may change the nature of the transition in the system. Note that in the limit $p \to \infty$, our model becomes equivalent to the nearest-neighbor QCP in previous works~\cite{marcuzzi2,buchhold}.

\subsection{Total Hamiltonian}
\label{sec:Hamiltonian}
\label{sec:2}

By solving the Lindblad equation, Eq.~\eqref{eq:lindqcp}, one may find the phase diagram of the system; however, this
is not easy when system size $N$ becomes large, $N \gg 1$. Instead, in this work we take the semiclassical approach starting with the quantum Langevin equation, as seen in previous works for the nearest-neighbor QCP~\cite{marcuzzi2, buchhold}.

To derive the Langevin equations, we first set the equivalent Hamiltonian for a system with $N$ spins, $N_b$ harmonic baths, and their interactions, where $N_b$ is given by $N_b = 2N^2-N$, as follows. The total Hamiltonian should be given by
\begin{equation}
\hat{H}_{\rm tot} = \hat{H}_S+ \sum_{l} \hat{H}_{d}(l)
+\sum_{m,l} \left[ \hat{H}_{b}(m,l)+\hat{H}_{c}(m,l) \right] \,,
\label{eq:H}
\end{equation}
where $\hat{H}_S$ is the same as in Eq.~\eqref{eq:HS}. There are three types of Hamiltonians for the baths and interactions (see Fig.~1). First, the Hamiltonian $\hat{H}_d$, which corresponds to the decay process, is assigned to each spin, and $\hat{H}_d(l)$ defined on spin $l$ is then given by
\begin{equation}
\hat{H}_{d}(l) = \sum_{q} \theta_q\hat{d}^{\dagger}_{l,q}\hat{d}_{l,q}+
\sum_{q}\left[ \lambda_{q}
\hat{d}^{\dagger}_{l,q} \hat{\sigma}^{-}_{l}+h.c.\right] \,,
\label{eq:Hd}
\end{equation}
where $\theta_q$ denotes the energy of bath particles with momentum $q$, and $h.c.$ stands for the Hermitian conjugate.
Here, $\hat{d}^{\dagger}_{l,q}$ and $\hat{d}_{l,q}$ are the creation and annihilation operators, respectively, of
particles of the bath associated with spin $l$, and $\lambda_q$ is the coupling strength of the decay process of the active state accompanied by emission of a single bath particle. Note that baths having different site indices are mutually independent, which is represented in the commutation relation for $\hat{d}^{\dagger}_{l,q}$ and $\hat{d}_{l,q}$, that is, $[\hat{d}_{l,q},\hat{d}_{m,q'}^{\dagger}]=\delta_{l,m}\delta_{q,q'}$.

The other Hamiltonians, for branching ($\hat{H}_b$) and coagulation ($\hat{H}_c$), are defined at each link
$(l,m)$ with direction, which means $\hat{H}_{b(c)}(m,l) \neq \hat{H}_{b(c)}(l,m)$. The branching and coagulation Hamiltonians are also given by the bath energy and interactions, similar to that of the decay process. Because branching and coagulation are allowed between long-distance spins, the interaction between the system and bath particles contains the distribution $P(|{\bf x}_m - {\bf x}_l|)$, so the Hamiltonians are given by
\begin{eqnarray}
\label{eq:Hb}
\hat{H}_{b}(m,l) &=& \sum_{q} \phi_q\hat{b}^{\dagger}_{ml,q}\hat{b}_{ml,q}+\sum_q \left[\chi_q
\sqrt{P_{ml}}
\,\hat{b}^{\dagger}_{ml,q}\, \hat{n}_m\hat{\sigma}^+_l+h.c.\right]\,, \nonumber \\ \\
\label{eq:Hc}
\hat{H}_{c}(m,l) &=& \sum_{q} \phi_q\hat{c}^{\dagger}_{ml,q}\hat{c}_{ml,q}+\sum_{q}
\left[\chi_q \sqrt{P_{ml}}
\,\hat{c}^{\dagger}_{ml,q}\, \hat{n}_m\hat{\sigma}^-_l+h.c. \right]\,, \nonumber\\ 
\end{eqnarray}
where $P_{ml}$ is shorthand notation for the L{\'e}vy distribution. Further, 
$\hat{b}_{ml,q}$ and $\hat{c}_{ml,q}$ are also operators of the harmonic baths defined
on the directional link, which
satisfy the commutation relation, $[\hat{b}_{ml,q},\hat{b}_{m'l',q}^{\dagger}]
=\delta_{l,l'}\delta_{m,m'}$ (the relations for $\hat{c}_{lm,q}$ and
$\hat{c}_{lm,q}^{\dagger}$ are obtained by
replacing $\hat{b}$ and $\hat{b}^{\dagger}$ with $\hat{c}$ and $\hat{c}^{\dagger}$,
respectively). Further, $\phi_q$ and
$\chi_q$ are the energy function and coupling between the system and baths, respectively, where
the branching and coagulation Hamiltonians share the same functions. 
Note that we have thus used $N_b$ independent baths, including 
$N$ decay, $N(N-1)$ branching, and $N(N-1)$ coagulation baths.

To obtain Eq.~\eqref{eq:lindqcp} for the density operator,
we consider that the bath is in a pure state $| 0\rangle_{B}$ with zero temperature
such that $ \hat{d} | 0\rangle_{B} = \hat{b} | 0\rangle_{B} = \hat{c} | 0\rangle_{B} = 0$.
Following the Born--Markov approximation, where the density operator of the total
system can be given approximately by the product state, $\hat{\rho}(t)
\otimes \hat{\rho}_B$, with the stationary bath density
$\hat{\rho}_B= |0\rangle \langle 0|_{B}$~\cite{breuer},
the density operator $\hat{\rho}(t+dt)$ is given by
\begin{equation}
\hat{\rho}(t+dt) = {\rm tr}_B \left \{
e^{-i dt \hat{H}_{\rm tot}}
\hat{\rho}(t) \otimes |0\rangle \langle 0|_{B}
e^{i dt \hat{H}_{\rm tot}} \right \} \,.
\label{eq:lind_derv}
\end{equation}
Expanding the evolution operators up to the second order of $dt$
and using $\langle \hat{H}_{\rm I} \rangle_B$, where
$ \hat{H}_{\rm I} $ is the part of Eq.~\eqref{eq:H} representing the interaction between the system and bath, and $\langle \cdot \rangle_B$
denotes $\langle 0| \cdot |0 \rangle_B$ with the bath state,
we write Eq.~\eqref{eq:lind_derv} up to the order of $dt^2$:
\begin{eqnarray}
\label{eq:lind_derv2}
\hat{\rho}(t+dt) &=& \hat{\rho}(t)  -i dt \left[ \hat{H}_S ,
\hat{\rho}(t) \right] \\ &&+ dt^2
\left( \hat{H}_S \hat{\rho}(t) \hat{H}_S
-\frac{1}{2}\left\{ \hat{H}_S^2 , \hat{\rho}(t)  \right\} \right)
\nonumber\\
&&+ dt^2
\left( \langle \hat{H}_{\rm I}\, \hat{\rho}(t) \otimes
|0\rangle \langle 0|_B  \hat{H}_{\rm I} \rangle_B
-\frac{1}{2}\left\{ \langle \hat{H}_{\rm I}^2 \rangle_B, \hat{\rho}(t)  \right\} \right) \,. \nonumber
\end{eqnarray}
Because $\langle \hat{d}_q\hat{d}^{\dagger}_q \rangle_B=1$, 
$\langle \hat{b}_q\hat{b}^{\dagger}_q \rangle_B=1$, and
$\langle \hat{c}_q\hat{c}^{\dagger}_q \rangle_B=1$, 
and otherwise the correlators of bath particles are zero,
the third line in Eq.~\eqref{eq:lind_derv2} can be written as
\begin{eqnarray} 
&&~dt^2 \sum_{l}  \sum_q \lambda_q^2
\left[ \hat{\sigma}^{-}_{l}\hat{\rho} \left( \hat{\sigma}^{-}_{l} \right)^{\dagger}
-\frac{1}{2} \left\{ \left( \hat{\sigma}^{-}_{l} \right)^{\dagger}
\hat{\sigma}^{-}_{l},\hat{\rho} \right\} \right] \nonumber\\
&&+ dt^2 \sum_{l,m} \sum_q \chi_q^2
P_{ml} \left[ \hat{n}_m\hat{\sigma}^+_l
\hat{\rho} \left(\hat{n}_m\hat{\sigma}^+_l\right)^{\dagger}
-\frac{1}{2} \left\{\left(\hat{n}_m\hat{\sigma}^+_l\right)^{\dagger}
\hat{n}_m\hat{\sigma}^+_l ,\hat{\rho} \right\} \right]\nonumber\\
&&+ dt^2 \sum_{l,m} \sum_q \chi_q^2
P_{ml}\left[ \hat{n}_m\hat{\sigma}^-_l
\hat{\rho} \left(\hat{n}_m\hat{\sigma}^-_l\right)^{\dagger}
-\frac{1}{2} \left\{\left(\hat{n}_m\hat{\sigma}^-_l\right)^{\dagger}
\hat{n}_m\hat{\sigma}^-_l ,\hat{\rho} \right\} \right].\nonumber
\end{eqnarray}
In accordance with the Weisskopf--Wigner theory~\cite{scully},
we extract the slow mode of bath particles around $q=0$ by setting 
$\lambda_q \approx \lambda_{q=0} $ 
and $\chi_q \approx \chi_{q=0}$. 
Then, the summations over $q$
become $\sum_{q} \lambda^2_q \approx \lambda_0^2 \sum_q$ and
$\sum_{q} \chi^2_q \approx \chi_0^2 \sum_q$.
To evaluate $\sum_{q}$, we use
the definition of the Dirac delta function, 
$(2\pi)^{-1} \sum_{q} \exp(-i\omega_q \tau) = \delta(\tau)$, with a linear function $\omega_q$ of $q$.
Inserting $\tau=0$ in both $\exp(-i\omega_q \tau)$ and $\delta(\tau)$, one can see       
\begin{equation}
dt^2 \,\lambda^2_0 \sum_{q} \to dt\, 2\pi \lambda^2_0\,,\quad
dt^2\, \chi^2_0 \sum_{q} \to dt\, 2\pi \chi^2_0\,, \nonumber
\end{equation} 
where we have used the fact that $\delta(0) \to 1/dt$ as $dt \to 0$.
Therefore, we can reduce $dt^2$ to $dt$ in the third line of Eq.~\eqref{eq:lind_derv2}.
Defining $\gamma = 2 \pi \lambda_0^2$ and $\kappa = 2 \pi \chi_0^2$, and retaining the 
order of $dt$,
we arrive at the Lindblad equation [Eq.~\eqref{eq:lindqcp}] from Eq.~\eqref{eq:lind_derv2}.

\subsection{Quantum Langevin equation}
\label{sec:Langevin}
Now, we derive the equations of motion for the system degrees of freedom 
from the Heisenberg equation in the total Hilbert space.
During the procedure, noise and the influence of heat baths will be defined so that
the quantum Langevin equation, which is the starting point for
the semiclassical theory of the long-range QCP, can be obtained.

For the system operators $\hat{a}_l=\hat{\sigma}_l^x,\hat{\sigma_l}^y,\hat{n}_l$, which
are Hermitian operators,
the Heisenberg equation is given by $\partial_t\hat{a}_l(t) = i[\hat{H},\hat{a}_l(t)]$; then
\begin{equation}
\label{eq:heisen}
\partial_t\hat{a}_l = i \left\{ \left[ \hat{H}_S,\hat{a}_l\right] +
 \left[\hat{H}_{d}(l),\hat{a}_l\right] 
+\sum_{m,j=b,c} 
\left[\hat{H}_{j}(m,l)+ \hat{H}_{j}(l,m), \hat{a}_l\right]\right\} \,.
\end{equation}
The first term in Eq.~\eqref{eq:heisen} consists only of system operators:
\begin{equation}
i\left[ \hat{H}_S,\hat{a}_l\right] = i \omega \sum_{m} P_{ml} \left(
\left[ \hat{n}_{m} \hat{\sigma}^+_l , \hat{a}_l \right]
+ \left[ \hat{n}_l \hat{\sigma}^+_{m} , \hat{a}_l \right]
+ h.c. \right) \,,
\label{eq:heisen1}
\end{equation}
and the other terms,
which are obtained from commutation with the interaction terms,
are mixtures of the system and bath operators, as shown below.
The commutation with the decay Hamiltonian reads
\begin{equation}
\label{eq:heisen2}
i\left[ \hat{H}_{d}(l),\hat{a}_l \right] =   \sum_{q} \lambda_q \left( i\hat{d}^{\dagger}_{l,q}
\left[ \hat{\sigma}^-_{l},\hat{a}_l\right]
-i\left[\hat{a}_l, \hat{\sigma}^+_{l}\right]\hat{d}_{l,q} \right)\,,
\end{equation}
that with the branching Hamiltonian is 
\begin{equation}
\label{eq:heisen3}
i\left[ \hat{H}_{b}(m,l) ,\hat{a}_l \right] =
 \sum_q  \chi_q \sqrt{P_{ml}}  \,\left(i \hat{b}^{\dagger}_{ml,q}
\left[\hat{n}_m \hat{\sigma}^+_{l},\hat{a}_l \right] 
+h.c. \right)\,,  
\end{equation}
and finally, that with the coagulation Hamiltonian is 
\begin{equation}
\label{eq:heisen4}
i\left[ \hat{H}_{c}(m,l) ,\hat{a}_l \right] =
 \sum_q  \chi_q \sqrt{P_{ml}}  \,\left( i\hat{c}^{\dagger}_{ml,q}
\left[\hat{n}_m \hat{\sigma}^-_{l},\hat{a}_l \right] 
+ h.c. \right)\,.  
\end{equation}
Note that $i[\hat{H}_{b(c)}(l,m) ,\hat{a}_l]$ can be obtained by replacing $l$ and $m$ with each other, except for $\hat{a}_l$,
in Eqs.~\eqref{eq:heisen3} and \eqref{eq:heisen4}.
Here, the Heisenberg picture has been used for all operators such that $\hat{a}_{l} \equiv \hat{a}_{l}(t)=
e^{i\hat{H}t } \hat{a}_{l}(0) e^{-i \hat{H} t}$, where $\hat{a}_{l}(0)$ denotes the Schr{\"o}dinger operator.
Henceforth, an operator without an explicit time represents the Heisenberg operator at time $t$.

To proceed a step further, we need functional forms of $\hat{b}_{ml,q}$, $\hat{c}_{ml,q}$,
and $\hat{d}_{l,q}$, whose equations of motion are also obtained from the Heisenberg equations,
$\partial_t\hat{d}_{l,q} = i [\hat{H}_{d}(l), \hat{d}_{l,q} ]$,
$\partial_t\hat{b}_{ml,q} = i [\hat{H}_{b}(m,l), \hat{b}_{ml,q} ]$, and
$\partial_t\hat{c}_{ml,q} = i [ \hat{H}_{c}(m,l) ,\hat{c}_{ml,q}  ]$, respectively.
It is easy to show that the solutions are given by
\begin{eqnarray}
\label{eq:sold}
\hat{d}_{l,q} &=& \hat{d}_{l,q}(0)e^{-i\theta_qt}-i\lambda_q\int_0^td\tau\,\hat{\sigma}_l^-(\tau) 
e^{-i\theta_q(t-\tau)}\,, \\
\label{eq:solb}
\hat{b}_{ml,q} &=& \hat{b}_{ml,q}(0)e^{-i\phi_qt}\nonumber \\
&&~~-i\chi_q
\sqrt{P_{ml}}\int_0^td\tau\,\hat{n}_m(\tau)\hat{\sigma}_l^+(\tau)e^{-i\phi_q(t-\tau)}\,, \\
\label{eq:solc}
\hat{c}_{ml,q} &=& \hat{c}_{ml,q}(0)e^{-i\phi_qt}\nonumber\\
&&~~-i\chi_q
\sqrt{P_{ml}}\int_0^td\tau\,\hat{n}_m(\tau)\hat{\sigma}_l^-(\tau)e^{-i\phi_q(t-\tau)}\,.
\end{eqnarray}
Plugging Eqs.~\eqref{eq:sold}--\eqref{eq:solc} into Eqs.~\eqref{eq:heisen2}--\eqref{eq:heisen4}
reveals that
the equation of motion for $\hat{a}_{l}$ can be divided into two parts, where one part is composed only of
the system operators, and the other contains both system and bath operators.
For example, inserting Eq.~\eqref{eq:sold} into Eq.~\eqref{eq:heisen2}, we get
\begin{eqnarray}
\label{eq:heisen22}
i\left[ \hat{H}_{d}(l),\hat{a}_l \right] &=&   \sum_{q} \lambda_q\left( i\hat{d}^{\dagger}_{l,q}(0)
\left[ \hat{\sigma}^-_{l},\hat{a}_l\right]e^{i\theta_q t}
+h.c. \right)    \\
&&- \sum_{q} \lambda_q^2\int_0^t d\tau \left(  \hat{\sigma}^+_{l}(\tau)
\left[ \hat{\sigma}^-_{l},\hat{a}_l\right]e^{i\theta_q (t-\tau) }
+ h.c. \right)\,, \nonumber
\end{eqnarray}
where the first line gives the quantum noise from the bath, and the 
second line is the dissipative term. Employing the Weisskopf-Wigner theory 
as shown in the previous section~(\ref{sec:Hamiltonian}), we let 
$\lambda_q$ be constant in Eq.~\eqref{eq:heisen22}, which leads to
$\lambda_0^2\sum_q  e^{i\theta_q (t-\tau)} 
\approx 2 \pi \lambda_0^2 \delta (t-\tau)$.
Therefore, Eq.~\eqref{eq:heisen22} reads
\begin{eqnarray}
\label{eq:heisen23}
&& \sqrt{\frac{\gamma}{2\pi}}\sum_{q} \left( i\hat{d}^{\dagger}_{l,q}(0)e^{i\theta_q t}
\left[ \hat{\sigma}^-_{l},\hat{a}_l\right]
- i\left[ \hat{a}_l, \hat{\sigma}^+_{l}\right]\hat{d}_{l,q}(0)e^{-i\theta_q t}\right) \nonumber\\
&&~~~~- \frac{\gamma}{2} \left(  \hat{\sigma}^+_{l}
\left[ \hat{\sigma}^-_{l},\hat{a}_l\right]
+ \left[\hat{a}_l, \hat{\sigma}^+_{l}\right]\hat{\sigma}^-_{l} \right)\,,
\end{eqnarray}
because $\int_0^t d\tau \delta(t-\tau) = 1/2$, and $\gamma = 2 \pi \lambda_0^2$.
For Eqs.~\eqref{eq:heisen3} and
\eqref{eq:heisen4}, one can also obtain similar expressions with 
$\kappa = 2 \pi \chi_0^2$. 

Because the second line in Eq.~\eqref{eq:heisen23} can be rewritten as 
\begin{equation}
\left(\sqrt{\gamma} \hat{\sigma}^-_{l}\right)^{\dagger} \hat{a}_l 
\sqrt{\gamma} \hat{\sigma}^-_{l}  -\frac{1}{2}\left\{ \left(\sqrt{\gamma} \hat{\sigma}^-_{l}\right)^{\dagger}
\sqrt{\gamma} \hat{\sigma}^-_{l}, \hat{a}_l \right \} 
\equiv\hat{F}(\hat{a}_l,\sqrt{\gamma} \hat{\sigma}^-_{l}), \nonumber
\end{equation}
by combining Eqs.~\eqref{eq:heisen1}--\eqref{eq:heisen4} 
with the Weisskopf--Wigner theory,    
we obtain the equations of motion for $\hat{a}_l$, which are given by 
\begin{eqnarray}
\label{eq:qlangevin}
\partial_t \hat{a}_l &=&
 i \omega \sum_{m} P_{ml} \left(
\left[ \hat{n}_{m} \hat{\sigma}^+_l , \hat{a}_l \right]
+ \left[ \hat{n}_l \hat{\sigma}^+_{m} , \hat{a}_l \right]
+ h.c. \right)  \nonumber\\
&&+\sum_{\alpha,m}  \hat{F}\left( \hat{a}_l, \hat{I}^{\alpha}_{lm} \right)
  + \hat{\eta}(\hat{a}_l)\,,
\end{eqnarray}
where for convenience we defined the interaction operators 
$\hat{I}^{\alpha}_{lm}$ as
\begin{eqnarray}
\hat{I}^{1}_{lm} &=& \sqrt{\gamma}\, \hat{\sigma}^-_{l} \delta_{l,m}\,, \nonumber\\
\hat{I}^{2}_{lm} &=& \sqrt{\kappa P_{ml}}\, \hat{n}_m \hat{\sigma}^+_{l} \,,\,\,\, 
\hat{I}^{3}_{lm} = \sqrt{\kappa P_{ml}}\, \hat{n}_l \hat{\sigma}^+_{m} \,, \nonumber\\
\hat{I}^{4}_{lm} &=& \sqrt{\kappa P_{ml}}\, \hat{n}_m \hat{\sigma}^-_{l} \,,\,\,\,  
\hat{I}^{5}_{lm} = \sqrt{\kappa P_{ml}}\, \hat{n}_l \hat{\sigma}^-_{m} \,.
\end{eqnarray} 
The Kronecker delta function $\delta_{lm} =1$ for $l=m$ and is zero otherwise. 
Although $\hat{I}^{\alpha}_{lm}$ has the same form as the Lindblad jump operators in
Eqs.~\eqref{eq:ld}--\eqref{eq:lc}, it is composed of the Heisenberg operators defined at time $t$,
which are different from the Lindblad operators. Finally, the \emph{noise} operator $\hat{\eta}(\hat{a}_l,t)$
is also written in terms of $\hat{I}^{\alpha}_{lm}$ and the corresponding bath operators.
We redefine the bath operators $\hat{B}^{\alpha}_{lm,q}(t)$ with the original operators as 
\begin{eqnarray}
\hat{B}^1_{lm,q}(t) &=& \hat{d}_{l,q}(0)\,\delta_{l,m} e^{-i\theta_q t} \,,\nonumber\\
\hat{B}^2_{lm,q}(t) &=& \hat{b}_{ml,q}(0) e^{-i\phi_q t} \,,\,\,\, 
\hat{B}^3_{lm,q}(t) = \hat{b}_{lm,q}(0) e^{-i\phi_q t}\,, \nonumber\\
\hat{B}^4_{lm,q}(t) &=& \hat{c}_{ml,q}(0)e^{-i\phi_q t} \,,\,\,\, 
\hat{B}^5_{lm,q}(t) = \hat{c}_{lm,q}(0) e^{-i\phi_q t}\,.
\end{eqnarray}
Then,
one can write the noise operator in the compact form
\begin{equation}
\label{eq:qnoise}
\hat{\eta}(\hat{a}_l) = \frac{i}{\sqrt{2\pi}} 
\sum_{\alpha}\sum_{m,q}  \left(\hat{B}^\alpha_{lm,q} (t)\right)^{\dagger}
\left[ \hat{I}^{\alpha}_{lm} , \hat{a}_l \right] + h.c.
\end{equation}
Obviously, the quantum average of the noise operators becomes zero:
$\langle \hat{\eta}(\hat{a}_{l}) \rangle_B = 0$.

\section{Mean-field result}
\label{sec:3}
\subsection{Mean-field equations}
\label{sec:MF_eq}
To explore the MF phase transition of the QCP, we extract the MF equation from 
the quantum Langevin equation, Eq.~\eqref{eq:qlangevin},  
by taking the trace of the equations of the operators with
the initial density operator given by $\hat{\rho}(0)\otimes \hat{\rho}_B$. 
By defining 
\begin{equation} \label{eq:expectation}
a_l(t) \equiv \langle \hat{a}_l(t) \rangle= {\rm tr} \, \hat{a}_l(t) \hat{\rho}(0)\otimes \hat{\rho}_B \,, \end{equation} 
the equations of the fields can be obtained; for example, the equation of motion for $n_l$ is given by
\begin{equation}
\label{eq:field_n}
\dot{n}_l = \omega \sum_{m} P_{ml} \langle \hat{n}_m\hat{\sigma}_l^y \rangle
-\gamma n_l+ \kappa \sum_{m}P_{ml}\left( n_m
-2\langle \hat{n}_m \hat{n}_l \rangle \right) \,.
\end{equation}
One can also derive similar equations for $\sigma^x_l(t)$ and $\sigma^y_l(t)$.
Ignoring correlations such as $\langle \hat{n}_m\hat{\sigma}_l^y \rangle 
\to n_m(t) \sigma^y_l(t)$ and taking uniform fields, $n_l(t) \to n(t)$, 
$\sigma^x_l(t) \to \sigma^x(t)$, and $\sigma^y_l(t) \to \sigma^y(t)$,
we arrive at the MF equations, which are given by
\begin{eqnarray}
\label{eq:mf}
\dot{n} &=& \omega n \sigma^y
+(\kappa -1) n -2\kappa n^2 \,, \nonumber\\
\dot{\sigma}^x &=& -\omega \sigma^x\sigma^y
-\frac{1+\kappa }{2}\sigma^x -\kappa n\sigma^x \,,\nonumber \\
\dot{\sigma}^y &=& \omega \left\{ 2n
+\left(\sigma^x\right)^2- 4n^2 \right\}
-\frac{1+\kappa}{2}\sigma^y-\kappa n\sigma^y \,,
\end{eqnarray}
where we rescale time, $t\gamma \to t$, $\omega /\gamma \to \omega$, and
$\kappa/\gamma \to \kappa$.  
Note that the above equations are equivalent to the MF equation used in 
previous studies for the nearest-neighbor QCP model~\cite{marcuzzi2, buchhold}.

\subsection{Phase diagram}
\label{sec:MF_phase}
\begin{figure}
\includegraphics[width=0.9\columnwidth]{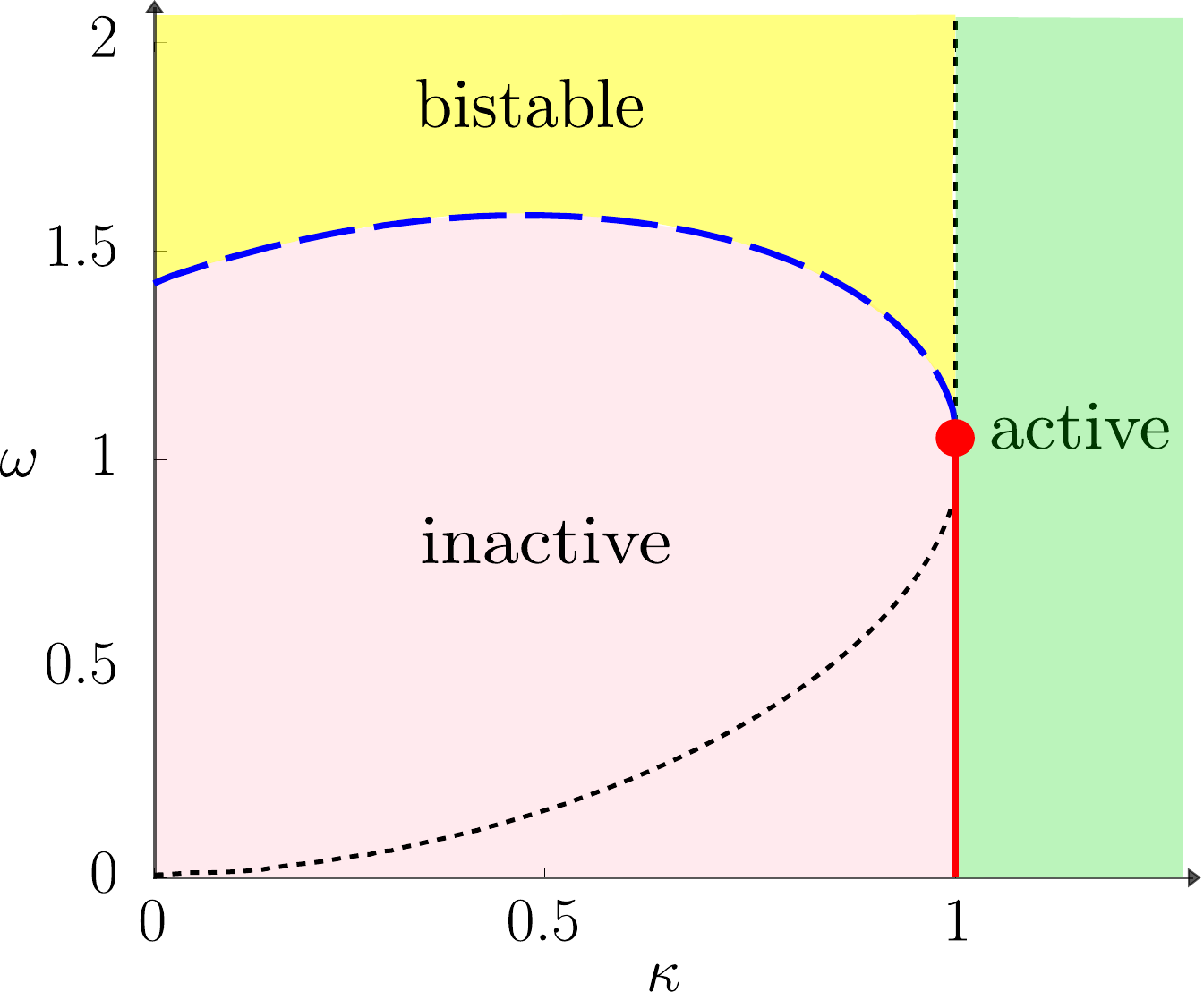}
\caption{Phase diagram of QCP. This diagram is represented as a plot of the classical rate $\kappa$ and the quantum rate $\omega$. In the weak quantum regime, the second-order transition is observed [(red) solid vertical line and (red) filled circle; Eq.~\eqref{eq:critical}]. In contrast, in the strong quantum regime, the absorbing transition is found to be of the first-order type [upper dashed (blue) curve between bistable and inactive states; Eq.~\eqref{eq:firstorder}]. The (red) filled circle, at which the two transitions intersect, is the tricritical point. 
(Black) dotted vertical line represents the boundary of the number of the stable solutions and (black) lower dotted curve inside the inactive region represents the boundary between the existence and nonexistence of multiple solutions.
\label{fig2}}
\end{figure}

In this section, we review the previous MF result, 
which is also similar to those in previous studies of TDP~\cite{henkel, grassberger_conjecture, lubeck, 
grassberger3}. 
It is found that the steady-state solutions or fixed points, $n_0$, $\sigma_0^x$,
and $\sigma^y_0$, satisfying
$\dot{a} =0$ in Eq.~\eqref{eq:mf}, form two groups as follows.
One is given by
\begin{eqnarray}
\label{eq:sol1}
&&\sigma^x_0=0\,,\qquad\sigma^y_0=\frac{4\omega n_0(1-2n_0)}{1+\kappa+2\kappa n_0} \,, \\
&&n_0 = 0\,\,, 
\frac{\omega^2-\kappa\pm\sqrt{(\omega^2-\kappa)^2+(\kappa^2+2\omega^2)(\kappa^2-1)}}{4\omega^2+2\kappa^2}\,, \nonumber
\end{eqnarray} 
and the other is given by
\begin{eqnarray}
\label{eq:sol2}
&&\sigma^x_0= \pm \sqrt{ 4n_0^2 -2n_0 - \left(1+\kappa + 2 \kappa n_0 \right)^2/ (2\omega)^2 }\,, \\
&&\sigma^y_0 = -\frac{1+\kappa +2\kappa n_0}{2\omega}\,, \qquad
n_0 = 0\,\,, \frac{1}{6} - \frac{1}{2\kappa} \,. \nonumber 
\end{eqnarray}
Note that if only real solutions are required, the latter should be ruled out because
solutions $n_0$ do not give real values of $\sigma^x_0$ in Eq.~\eqref{eq:sol2}.  
Moreover, the nonzero solutions $n_0=n_0^{+} \equiv
\frac{\omega^2-\kappa + \sqrt{(\omega^2-\kappa)^2+(\kappa^2+2\omega^2)(\kappa^2-1)}}{4\omega^2+2\kappa^2}$ and $n_0=n_0^{-} \equiv
\frac{\omega^2-\kappa - \sqrt{(\omega^2-\kappa)^2+(\kappa^2+2\omega^2)(\kappa^2-1)}}{4\omega^2+2\kappa^2}$
in Eq.~\eqref{eq:sol1} does not exist when 
$(\omega^2-\kappa)^2<(\kappa^2+2\omega^2)(1-\kappa^2)$, which is inside the (blue) dashed curve and 
lower (black) dotted curve in Fig.~\ref{fig2}. 

Now, we check the stability of Eq.~\eqref{eq:sol1} through 
linearization of Eq.~\eqref{eq:mf} around the fixed points.
Inserting $n = n_0 + \delta n$, $\sigma^y = \sigma^y_0 + \delta \sigma^y$,  
and $\sigma^x = \sigma^x_0 + \delta \sigma^x$ into Eq.~\eqref{eq:mf},
and expanding up to the linear order of perturbations, we then obtain the linear equation 
$\dot{\delta{\bf a}} = {\bf M} \delta{\bf a}$, where 
\begin{equation}\label{eq:linear}
\delta{\bf a} = \left( \delta n, \delta \sigma^y, \delta \sigma^x \right)^{\sf T}\,,
\end{equation}
and the matrix ${\bf M}$ is
\begin{equation}
\label{eq:matrix}
{\bf M}=\left(\begin{array}{ccc} 
\omega\sigma_0^y-4\kappa n_0 +\kappa-1 & \omega n_0 & 0\\ 
-8\omega n_0 -\kappa \sigma^y_0 +2\omega & -\frac{2\kappa n_0 + \kappa +1 }{2} & 0\\ 
0 & 0&  -\frac{2\omega \sigma^y_0 +2\kappa n_0+ \kappa +1 }{2}
\end{array}\right).
\end{equation}
For $n_0=\sigma^y_0=\sigma^x_0=0$, all the eigenvalues of ${\bf M}$ are negative at $\kappa <1$,
meaning that the fixed point is stable, whereas one of the eigenvalues becomes positive when $\kappa >1$. 
Thus, $\kappa=1$ is the boundary for the fixed point, $n_0=\sigma^y_0=\sigma^x_0=0$.

For the nonzero solutions $n_0 =n^{+(-)}_0$ in Eq.~\eqref{eq:sol1}, 
by investigating the eigenvalues of Eq.~\eqref{eq:matrix},
one can note that $n_0=n^+_0$ and
$n_0=n^-_0$ are stable when $n^+_0>0$ and $n^-_0 <0$, respectively. 
More precisely, $n_0=n_0^+, \sigma^y_0=\sigma^y_0(n_0^+), \sigma^x_0=0$ is the stable fixed point
when $n_0^+ >0$ and $n^+_0 > \frac{1}{6} - \frac{1}{2\kappa}$. 
Because negative density, $n<0$, is not physically allowed, 
the fixed point $n_0=n_0^-<0$ should be ruled out in this analysis.    
To find the stable region of $n^+_0$, first we note that 
$n^{+}_0 \ge 0$ becomes marginal along two curves.
One is the (red) solid vertical line including a (red) filled circle in Fig.~\ref{fig2},
\begin{equation} \label{eq:critical}
 \kappa=1 \,, \omega \le 1 \,,
\end{equation}
where $n^{+}_0 = n_0 =0$, and the other is the dashed (blue) curve, given by
\begin{equation} \label{eq:firstorder}
\omega = \left( 1+\kappa -\kappa^2 + \sqrt{(1+\kappa -\kappa^2 )^2 -\kappa^4 } \right)^{1/2}
\,\,\, {\rm at } \,\,\,
\kappa \le 1\,,
\end{equation}
where $n^+_0=n^-_0 \ge 0$. $n_0^+$ is found to be stable outside of the region enclosed
by the two curves, Eqs.~\eqref{eq:critical} and \eqref{eq:firstorder}.
Thus, the stability analysis yields the phase diagram shown in Fig.~\ref{fig2}.
With the boundaries described by Eqs.~\eqref{eq:critical} and ~\eqref{eq:firstorder}, 
there are three regions: (i) the inactive phase, $n_0=0$, (ii) the active phase, $n_0=n^+_0$, 
divided by Eq.~\eqref{eq:critical} with
a single stable fixed point, and (iii) a 
bistable phase possessing two stable fixed points, $n_0=0$ and $n_0=n^+_0$, with
the boundary curves of Eq.~\eqref{eq:firstorder} and $\kappa=\kappa_c$ with $\omega>1$.
The solutions of Eq.~\eqref{eq:sol1}  
show that between (i) and (ii) there exist second-order phase transitions 
with the order parameter exponent $\beta=1$ for $\omega<1$ and $\beta=1/2$ at $\omega=1$.
Moreover, one can observe that the first-order transition may occur between (i) and (iii), 
implying that $(\kappa, \omega)= (\kappa_c, 1)$ where the boundaries meet at the tricritical point.

By substituting the expression for $\sigma^y_0$ in Eq.~\eqref{eq:sol1} into the
equation for $n$ [Eq.~\eqref{eq:mf}],  
one may expand the equation with small $n_0$ near the critical line Eq.~\eqref{eq:critical} as
\begin{equation}
\label{eq:mf2}
 \dot{n}=0= -u_2 n_0 - u_3 n_0^2 - u_4 n_0^3 + \mathcal{O}(n_0^4),
\end{equation}
where $u_2= (\kappa_c-\kappa)$, and 
$u_3$ and $u_4$ are given by
\begin{equation}
u_3 = \frac{2\kappa (1+\kappa) - 4\omega^2}{1+\kappa} \,, \qquad 
u_4 = \frac{8 \omega^2 (1+ 2\kappa) }{( 1+\kappa)^2}  \,, 
\end{equation}
respectively. Note that Eq.~\eqref{eq:mf2} implies 
an effective MF potential defined as
\begin{equation} \label{eq:umf}
U_{MF} =  \sum_{k=2} \frac{ u_k}{k} n^k \,,   
\end{equation}     
where $u_k$ is defined in Eq.~\eqref{eq:mf2}.
Then, the solution $n_0$ satisfying Eq.~\eqref{eq:mf2} is also 
the steady-state solution of the single effective equation of the order parameter, 
which is given by 
\begin{equation}
\label{eq:mf3}
\dot{n} = -\partial U_{MF} / \partial n.  
\end{equation}
By expanding $U_{MF}$ up to the fourth order, it is found that when $\omega <1$,
$u_3$ and $u_4$ are positive near $\kappa =\kappa_c$. Consequently, $n_0=0$ becomes 
unstable, and the stable fixed point is given by 
$n_0 \approx (\kappa -\kappa_c)/u_3$ at $\kappa >\kappa_c$; 
consequently, the DP critical exponent $\beta =1$. On the other hand, 
at $\omega=1$, it is found that $u_3=0$ at $\kappa=\kappa_c$ yields
a different universality, called the TDP class, where  
the fixed point is given by $n_0 \approx \sqrt{(\kappa -\kappa_c)/u_4} $ with $\beta=1/2$. 
We conclude that the effective single equation Eq.~\eqref{eq:mf3} 
well describes the critical behavior in the steady state,
which is consistent with the linear stability analysis based on the MF equations 
of all the system variables in Eq.~\eqref{eq:mf}. 

We also observe that the effective equation Eq.~\eqref{eq:mf3}
captures a change in the nature of the transition
at the tricritical point $(\kappa, \omega)=(\kappa_c,1)$. 
Expanding $U_{MF}$ up to the fourth order again, as shown in Eq.~\eqref{eq:mf2}, 
one can see that all the positive coefficients $u_k$ yield the single fixed point $n_0=0$, but 
given negative $u_3$, an additional positive and stable fixed point can exist as
\begin{equation}\label{eq:jump}
n_0 = \frac{-u_3  + \sqrt{ u_3^2 -4u_2u_4}}{2u_4}\,,
\end{equation}
where $u_3^2 \ge 4u_2u_4$ is also satisfied. Because we consider 
only the limits $\kappa \to \kappa_c$ and $\omega \to 1$,  
the discriminant $u_3^2 = 4u_2u_4$ may give the curve near the tricritical point,
\begin{equation} \label{eq:firstorder2}
\omega =  1+ \sqrt{3(\kappa_c-\kappa)/2}  \,,
\end{equation}     
which is also obtainable by expanding Eq.~\eqref{eq:firstorder} at $\kappa = \kappa_c$.
Of course, there is another solution, $\omega =  1- \sqrt{3(\kappa_c-\kappa)/2}$, but
it does not satisfy $u_3<0$.
If $\omega > 1+ \sqrt{3 (\kappa_c-\kappa) /2} $ at $\kappa < \kappa_c$,
there are two stable fixed points, $n_0=0$ and Eq.~\eqref{eq:jump}, which is consistent with
the previous discussion of the linear stability analysis. 
Because the fixed point of Eq.~\eqref{eq:jump} disappears abruptly, and $n_0=0$ 
becomes the only fixed point crossing the curve of Eq.~\eqref{eq:firstorder2} 
from right to left, one may observe the first-order phase transition in this regime.   
Although the analysis of the MF potential for the first-order transition is
valid near the tricritical point, the entire analysis of linear stability with
Eq.~\eqref{eq:linear} 
implies a first-order transition with the transition line described by Eq.~\eqref{eq:firstorder}.

\section{Scaling behavior}
\label{sec:4}
\subsection{Phenomenological equation}   
\label{sec:peq}
To investigate the low-dimensional QCP, one may add spatiotemporal fluctuations to 
the MF equations, Eq.~\eqref{eq:mf}, as seen in previous works~\cite{marcuzzi2, buchhold},
where the action for $n$, $\sigma^x$, and $\sigma^y$ is obtained by the so-called 
Martin--Siggia--Rose--Janssen--de Dominicis (MSRJD) field theory~\cite{MSRJD1, MSRJD2, MSRJD3, MSRJD4, MSRJD5}.  
In this work, instead we start with
the effective MF equation of $n$, Eq.~\eqref{eq:mf3}, which is a plausible 
assumption near the critical line in Eq.~\eqref{eq:critical} 
because $\sigma^x$ and $\sigma^y$ may arrive quickly in the steady state [Eq.~\eqref{eq:sol1}]
owing to the finite gap energy, as seen in Eq.~\eqref{eq:mf}.
Near the critical line, by plugging $n_0 \approx \sigma^y_0 \approx 0$ and $\sigma^x_0=0$ into
Eq.~\eqref{eq:mf}, one can see that the excitation gap for $\sigma^x$ and $\sigma^y$ is
given by $(1+\kappa_c)/2$. 
Then, the critical dynamics can be described by a single equation associated with $n$ and based on
Eq.~\eqref{eq:mf3} with fluctuations. Using the standard MSRJD theory 
and the scaling theory,
we will show the critical exponents and upper critical dimensions of the long-range QCP. 

The phenomenological Langevin description has been regarded as a very useful method to 
study the critical phenomena of DP-type models~\cite{hinrichsen, henkel}, 
where the strength of the white noise is
proportional to the density of active states because
stochasticity is induced by the active states. 
We also follow the phenomenological approach to obtain the Langevin-type effective equation.  
Note that the field $n$ in the MF equations 
is the expectation value of the operator obtained by the trace over the density operator,
as seen in Eq.~\eqref{eq:expectation}, implying that
$n$ can also be thought of as an averaged field over the quantum noise 
manifested in the noise operator in Eq.~\eqref{eq:qlangevin}.
To describe the noise, one may start from the equations of operators. 
Instead of using the quantum noise operators [Eq.~\eqref{eq:qnoise}]
directly, we introduce a stochastic density field $\xi$ satisfying 
$\overline{ \xi } = n$, where
the overbar denotes the average over a phenomenological noise $\eta$.
We regard $\xi$ as a coarse-grained field of the active sites measured in a single realization, and take $\eta$ as the white Gaussian noise, which is plausible in thermodynamic limit. Near the critical point, $\xi$ may be governed by the effective Langevin equation keeping
the low energy fluctuations, given by
\begin{equation} \label{eq:xi}
\partial_t \xi_l = \kappa \sum_{m} P_{ml} \xi_m -\kappa \xi_l 
-\frac{\partial U_{MF}(\xi_l) }{\partial \xi_l} +\eta_l \,,
\end{equation}    
where the first term is the lowest order contribution of L{\'e}vy flight, and the potential $U_{MF}(\xi)$ 
is defined as having the same form as in Eq.~\eqref{eq:umf},
\begin{equation} 
U_{MF}(\xi) = \sum_{k=2} \frac{ u_k}{k} \xi^k \,.   
\end{equation}     
Setting $\overline{\eta } =0$, taking the average of $\eta$ in Eq.~\eqref{eq:xi},
and ignoring correlations such that $\overline{ \xi^k } \approx n^k$ with $k \geq 2$
and fluctuations,
one can obtain the same MF equation as Eq.~\eqref{eq:mf3} from 
the equation for $\xi$, Eq.~\eqref{eq:xi}.  

The noise $\eta$ should be invoked by the original quantum dynamics so that
Eq.~\eqref{eq:xi} reflects the original dynamics of $\hat{n}$. In the original dynamics, 
existing active states can generate stochastic processes such as decay and branching via
interactions with the baths. Therefore, we require that the strength $\mathcal{D}$ defined in 
\begin{equation} \label{eq:classicalD}
\overline{ \eta_{l}(t) \eta_{m}(t') } = \mathcal {D}_l\, \delta_{m,l}\, \delta(t-t') \,
\end{equation}
depends on the density $\xi_l$ as $\mathcal{D}_l \propto \xi_l$, implying
also that when $\xi=0$, there is no fluctuation, so the absorbing state is achieved.
Moreover, one may suspect that the original quantum noise itself 
also obeys the similar relation
\begin{equation} \label{eq:quantumD}
\langle \hat{\eta}(\hat{n}_l(t))\, \hat{\eta}(\hat{n}'_{m}(t')) \rangle_B \approx
\hat{\mathcal{D}}_l\, \delta_{m,l} \, \delta(t-t') \,,
\end{equation}
with $\hat{\mathcal{D}}_l \propto \langle \hat{n}_l \rangle_B$, where
$\hat{\mathcal{D}}$ is the strength of the quantum noise~\cite{scully, louisell}.  
Indeed, it has been revealed that the quantum noise strength in the nearest-neighbor QCP
is proportional to $\langle \hat{n} \rangle_B$~\cite{marcuzzi2, buchhold}.
If this is also true in our case, we can assume that the phenomenological noise $\eta$ originates from
the quantum noise operator $\hat{\eta}$ with a strength 
$\overline{ \mathcal{D}} = \langle \hat{\mathcal{D}} \rangle$, at least up to the leading order.
   
Now, we check the strength of the quantum noise in the long-range QCP, which is given by the
correlators of the noise operators in Eq.~\eqref{eq:qnoise}:
\begin{eqnarray}
\label{eq:cor}
\langle \hat{\eta}(\hat{a}_l(t))\, \hat{\eta}(\hat{a}'_{k}(t')) \rangle_B &=&
{\rm tr}_B \frac{1}{2\pi} \sum_{\alpha, \beta} \sum_{m,q}\sum_{m' ,q'} \left[ \hat{a}_l(t), 
\left(\hat{I}^{\alpha}_{lm}(t) \right)^\dagger \right]  \\
&&\times \hat{B}^\alpha_{lm,q} (t) \left(\hat{B}^\beta_{km',q'} (t')\right)^{\dagger}
\left[ \hat{I}^{\beta}_{km'}(t') , \hat{a}'_k(t') \right] \hat{\rho}_B \,, \nonumber
\end{eqnarray}
where again $\hat{\rho}_B = |0\rangle \langle 0|_B$.
By using the commutation relations of $\hat{B}^{\alpha}$, 
Eq.~\eqref{eq:cor} can be divided into three parts:
\begin{equation}
\label{eq:cor2}
\hat{\mathcal{D}}_1 \delta_{l,k}\, \delta(t-t') + \hat{\mathcal{D}}_2  \delta(t-t') 
+ \hat{\mathcal{D}}_3 \delta_{t,t'}  \,,
\end{equation}
where the first term is given by
\begin{equation}
\hat{\mathcal{D}}_1 = {\rm tr}_B \sum_{\alpha} \sum_{m} \left[ \hat{a}_l,
\left(\hat{I}^{\alpha}_{lm} \right)^\dagger \right] 
\left[ \hat{I}^{\alpha}_{lm} , \hat{a}'_l \right] \hat{\rho}_B \,,
\end{equation}
the second term $\hat{\mathcal{D}}_2$ reads
\begin{eqnarray}
\hat{\mathcal{D}}_2 &=& {\rm tr}_B \left \{ \left[ \hat{a}_l,
\left(\hat{I}^{2}_{lk} \right)^\dagger \right] 
\left[ \hat{I}^{3}_{kl} , \hat{a}'_k \right] 
+\left[ \hat{a}_l,
\left(\hat{I}^{3}_{lk} \right)^\dagger \right]     
\left[ \hat{I}^{2}_{kl} , \hat{a}'_k \right] \right. \nonumber\\
&&+ \left. \left[ \hat{a}_l,
\left(\hat{I}^{4}_{lk} \right)^\dagger \right] 
\left[ \hat{I}^{5}_{kl} , \hat{a}'_k \right] 
+\left[ \hat{a}_l,
\left(\hat{I}^{5}_{lk} \right)^\dagger \right]     
\left[ \hat{I}^{4}_{kl} , \hat{a}'_k \right] \right \} \hat{\rho}_B\,,
\end{eqnarray}
and finally, 
\begin{equation}
\hat{\mathcal{D}}_3 = {\rm tr}_B \frac{1}{4}\sum_{\alpha, \beta} \sum_{m ,m'} \left[ \left[ \hat{a}_l,
\left(\hat{I}^{\alpha}_{lm} \right)^\dagger \right], 
\left(\hat{I}^{\beta}_{km'} \right)^\dagger \right] 
 \left[ \hat{I}^{\alpha}_{lm},  \left[ 
\hat{I}^{\beta}_{km'},  \hat{a}'_k  \right] 
 \right] \hat{\rho}_B \,,  
\end{equation}
where we omitted the site indices in $\hat{\mathcal{D}}_{1,2,3}$.
To obtain $\hat{\mathcal{D}}_3$, we used the fact 
that the system operators and bath operators commute when they
are at the same time, for instance, $\hat{n}_l(t) \hat{d}_{l,q}(t) = \hat{d}_{l,q}(t) \hat{n}_l(t)$. 
Further, using 
the solutions of the bath particles, Eqs.~\eqref{eq:sold}--\eqref{eq:solc}, with the Weisskopf--Wigner theory,  
one can obtain the above form of $\hat{\mathcal{D}}_3$. 

In Eq.~\eqref{eq:cor2}, the contribution of $\hat{\mathcal{D}}_3$ can be
ignored because $\delta(t-t') \gg \delta_{t,t'}$ at $t=t'$. Moreover,
$\hat{\mathcal{D}}_2$ contains the contributions of only the pair $(l,k)$, whereas
$l$-to-all coupling contributes to $\hat{\mathcal{D}}_1$. Therefore, 
the strength of the noise, 
including $\hat{\mathcal{D}}$ 
in Eq.~\eqref{eq:quantumD}, may be determined mainly by $\hat{\mathcal{D}}_1$. Because $\hat{\mathcal{D}}_1$ is not a Hermitian
operator, to obtain a real value, we take
\begin{equation} \label{eq:realqD}
{\rm Re} \langle \hat{\mathcal{D}}_1 \rangle \equiv 
\langle \hat{\mathcal{D}}_1 + \hat{\mathcal{D}}^{\dagger}_1 \rangle /2. 
\end{equation} 
Now, we can obtain the noise strength ${\rm Re} \langle \hat{\mathcal{D}}_l \rangle $ 
for $\hat{\eta}(\hat{n}_l)$ from Eq.~\eqref{eq:realqD} by setting 
$\hat{a}_l=\hat{a}'_l =\hat{n}_l$:
\begin{equation} \label{eq:realqD2}
{\rm Re} \langle \hat{\mathcal{D}_l} \rangle = n_l  + 
\kappa\sum_{m} P_{ml} n_m  \,.
\end{equation} 
Because we are interested in the critical dynamics, where long-wavelength excitation
is crucial, we use the approximation $n_m\approx n_l$ for all $m$ in the summation term
in Eq.~\eqref{eq:realqD2}, which leads to 
\begin{equation} \label{eq:realqD3}
{\rm Re} \langle \hat{\mathcal{D}_l} \rangle \approx (1+\kappa) n_l  \,. 
\end{equation} 
This is what we expected, and now we take $\mathcal{D}_l = (1+\kappa) \xi_l$ 
for the noise strength in our Langevin equation [Eq.~\eqref{eq:xi}].

We point out that the leading order of 
the noise strength for $\hat{\sigma}^x$ or $\hat{\sigma}^x$ 
is given by a constant; more precisely, 
\begin{equation}
{\rm Re}\, \langle \hat{\eta}(\hat{\sigma}^x_l(t) )\, \hat{\eta}(\hat{\sigma}^x_l(t') )  \rangle \approx 
{\rm Re}\, \langle \hat{\eta}(\hat{\sigma}^y_l(t) )\, \hat{\eta}(\hat{\sigma}^y_l(t'))  \rangle \approx 
(1+\kappa) \delta(t-t')\,. \nonumber
\end{equation}
Moreover, the noise operator $\hat{\eta}(\hat{n} )$ is correlated with 
$\hat{\eta}(\hat{\sigma}^x )$ and $\hat{\eta}(\hat{\sigma}^y )$ as follows:
\begin{eqnarray}
{\rm Re}\, \langle \hat{\eta}(\hat{n}_l(t) )\, \hat{\eta}(\hat{\sigma}^x_l(t') ) \rangle &\approx&
 \sigma_l^x \delta(t-t')/2 \,, \nonumber\\
{\rm Re}\, \langle \hat{\eta}(\hat{n}_l(t) )\, \hat{\eta}(\hat{\sigma}^y_l(t')) \rangle &\approx&
\sigma_l^y \delta(t-t')/2 \,. \nonumber 
\end{eqnarray}
Thus, even if there is no active state at some point, active states can be induced by fluctuations
of $\hat{\sigma^x}$ and $\hat{\sigma^y}$, which implies that the absorbing state cannot be achieved.
This is reminiscent of the quantum fluctuation induced by the uncertainty relations between
the Pauli spin operators. 
Therefore, our semiclassical approach must be associated with a proper time scale, where
the quantum fluctuation is negligible. At this stage, we assume the time scale without
proof. 

In short, we introduced the stochastic field $\xi$ as the density field of active states and its phenomenological Langevin equation. To capture the critical dynamics of QCP, we took the lowest-order fluctuation in the long-range interaction to the MF equation of the order parameter $n$. Since the original dynamics shows the absorbing transition, we assumed that the strength of the white Gaussian noise is proportional to the density field.
Indeed, we confirmed that the original quantum noise also has the multiplicative nature, so we adopted the functional form of the quantum-noise strength in the lowest order as one of our phenomenological noise $\eta$. Because the Langevin equation of $\xi$ is the classical field equation, one can apply the classical field theory
to the QCP effectively at least near the critical point.
Finally, we remark that the quantum Langevin equation can be
transformed to the c-number Langevin equation
~\cite{louisell,c-number}. One may apply the conversion method 
in this work and expect to obtain a similar equation to ours, Eq.~\eqref{eq:xi}.
To check whether our assumptions are adequate and 
resolve the problem of time scale, 
it is worth studying the relationship between the phenomenological 
and c-number Langevin equations.   

\subsection{Critical exponents and upper critical dimensions}
\label{sec:upc}
\begin{table*}
\caption{MF critical exponents.
These critical exponents are obtained using the scaling transformation of Eq.~\eqref{eq:action}.
The universality classes are determined by the power of the long-range interaction ($p$) and the strength of the coherent dynamics ($\omega$ in ${\hat H}_s$). The long-range interaction is relevant (irrelevant) for $p\leq 2$ ($p>2$). Depending on whether $u_3=0$ or $u_3 > 0$, $d_c$ and $\beta$ can vary. 
}
\begin{center}
\setlength{\tabcolsep}{12pt}
{\renewcommand{\arraystretch}{1.5}
\begin{tabular}{cccccccc}
    \hline
    \hline
    
    & & $d_c$ & $\beta$ & $\nu_{\bot}$ & $\nu_{\|}$ & $z$  \\
    \hline
    \multirow{2}{*}{$\kappa=\kappa_c\,, \omega=1$} & $p>2$\,(TDP) & $3$ & $1/2$ & $1/2$ & $1$ & $2$\\
    \cline{2-2}
   ($u_3=0$)& $p\leq 2$\,(long-range TDP) & $3p/2$ & $1/2$ & $1/p$ & $1$ & $p$  \\
    \hline
    \multirow{2}{*}{$\kappa=\kappa_c\,, \omega<1$}
    & $p>2$\,(DP) & $4$ & $1$  & $1/2$ & $1$ & $2$\\
    \cline{2-2}
     ($u_3>0$)& $p\leq 2$\,(long-range DP) & $2p$ & $1$ & $1/p$ & $1$ & $p$  \\
    \hline
    \hline
\end{tabular}}
\label{table}
\end{center}
\end{table*}

To apply the scaling theory, the equation for continuous fields is more convenient than
the discrete equation. Taking the continuum limit with an appropriate rescaling like 
$\partial_t \to \tau \partial_t$, where $\tau$ is a scaling parameter, and 
expanding the L{\'e}vy term up to two leading orders, as in previous 
works~\cite{LDP1, LDP2, LDP3, LDP4, LDP5, LDP6, montroll}, 
we write the Langevin equation of the continuous density field $\xi=\xi({\bf r},t)$ up to
the $u_4$ term as
\begin{equation}
\tau\partial_t \xi = D \nabla^2 \xi + D_p\nabla^{p} \xi
-u_2 \xi - u_3\xi^2 -u_4 \xi^3 +\eta \,.
\label{eq:xi2}
\end{equation}
Here $D$ and $D_p$ are the diffusion constants, obtained from the
expansion, given by $\kappa \int \,d {\bf r}' P(|{\bf r} - {\bf r}'|) \xi ({\bf r}') 
\approx  \kappa \xi + D \nabla^2 \xi + D_p\nabla^{p} \xi $, 
and the noise $\eta({\bf r}, t)$ in the continuum limit 
obeys
\begin{equation}
\overline{ \eta({\bf r}_1, t_1) \eta({\bf r}_0, t_0) } = \Gamma \xi({\bf r}_0, t_0) \, \delta({\bf r}_1 - {\bf r}_0) 
\,\delta(t_1-t_0) \,,
\end{equation}
where $\Gamma = (1+\kappa)$. Note that $u_k$ in Eq.~\eqref{eq:xi2} was also rescaled appropriately.

Setting $\Gamma=0$, which yields $\xi = n$, one can obtain the MF exponents for 
the correlation length, $\nu_{\bot}$, and time, $\nu_{\|} = z \nu_{\bot}$.
Under the scaling transformations, which are given by
\begin{equation} \label{eq:transform}
|{\bf r}| \to |{\bf r}'|= s|{\bf r}| \,,\,\, t \to t'=s^zt \,, \,\, 
\xi \to \xi' \,,  
\end{equation}
where $s>1$, the transformed equation is written as
\begin{equation}\label{eq:xi3}
\tau s^{-z} \partial_t \xi' = D s^{-2} \nabla^{2} \xi' + D_p s^{-p} \nabla^{p} \xi'
-u_2 \xi' - u_3\xi'^2 -u_4 \xi'^3 \,.
\end{equation}
Because near the critical point the order parameter 
obeys the scaling form of $\xi'({\bf r'}) = s^{-\beta/\nu_{\bot}} \xi({\bf r})$
~\cite{hinrichsen,henkel}, we rewrite Eq.~\eqref{eq:xi3}
in terms of $\xi$ as 
\begin{equation}\label{eq:xi4}
\tau\partial_t \xi = Ds^{z-2} \nabla^{2} \xi + D_p s^{z-p} \nabla^{p}\xi
-u_2 s^{z}\xi - u_3 s^{z-\beta/\nu_{\bot}} \xi^2 - u_4 s^{z-2\beta/\nu_{\bot}} \xi^3 . \nonumber
\end{equation}
When $p>2$ and $u_3>0$, one may set $z=2$ and $\beta/\nu_{\bot}=z=2$; then, 
at the critical point where $u_2=0$, the equation given by
\begin{equation}
\tau\partial_t \xi = D \nabla^{2} \xi- u_3 \xi^2 \, \nonumber
\end{equation}   
is invariant under the scaling transformation 
because $D_p s^{z-p} \nabla^{p} \xi$ and $u_4 s^{z-2\beta/\nu_{\bot}} \xi^3$ vanish by
repeated transformations. Using the value $\beta=1$,  
we obtain the exponents, $\nu_{\bot}= 1/2$, and thus
$\nu_{\|} = 1$. These exponents belong to the DP class.

For $p >2$ and $u_3=0$, however, the relevant equation is given by
\begin{equation} 
\tau\partial_t \xi = D \nabla^{2} \xi- u_4 \xi^3 \,, \nonumber
\end{equation}  
so $\beta/\nu_{\bot}=z/2=1$. Using $\beta=1/2$ at the tricritical point corresponding to $u_3=0$,
we obtain the exponents $\nu_{\bot}= 1/2$ and $\nu_{\|} = 1$, which correspond to the TDP universality.
Therefore, if $p>2$, the long-range term becomes irrelevant for both $u_3>0$ and
$u_3=0$, so the universality
is equal to that in the short-range model. 
On the other hand, if $p<2$, one can see that the relevant term becomes
$D_p s^{z-p} \nabla^{p} \xi$ instead of $D s^{z-2} \nabla^{2} \xi$, leading to
the dynamic exponent $z=p$. Consequently, $\beta/\nu_{\bot}=z=p$ for $u_3>0$, whereas $\beta/\nu_{\bot}=z/2=p/2$ for $u_3=0$, 
yielding $\nu_{\bot}=1/p$ for both cases. 
The MF exponents for the short-range and long-range cases 
are summarized in Table~\ref{table}.    

To check the relevance of the noise, we employ the path integral formalism including the noise term in Eq.~\eqref{eq:xi2}.
Using the MSRJD theory for the Langevin equation, we obtain the action 
$S=S[\xi, \tilde{\xi} ]$ for Eq.~\eqref{eq:xi2}, where $\tilde{\xi}$ is the response field, as follows:
\begin{equation} \label{eq:action}
S =\int d{\bf x} \, \tilde{\xi} \left[
\tau\partial_t  - D \nabla^2 - D_p\nabla^{p} +u_2 
+ u_3\xi + u_4 \xi^2 -\frac{\Gamma}{2} \tilde{\xi} \right] \xi \,,
\end{equation}
where ${\bf x} = ({\bf r}, t)$. Under the transformation given by Eq.~\eqref{eq:transform},
$S[\xi, \tilde{\xi}] \to S' [\xi', \tilde{\xi}']$, where $S'$ can be written in terms of $\xi$ and $\tilde{\xi}$
using the relations $\xi'= s^{-b}\xi$ and $\tilde{\xi}' = s^{-\tilde{b}} \tilde{\xi}$, and is given by
\begin{eqnarray}
S' &=&\int d{\bf x} \, s^{d+z} \tilde{\xi}  \left[
\tau s^{-z - b-\tilde{b}} \partial_t  - D s^{-2 - b-\tilde{b}} \nabla^2 
- D_p s^{-p - b-\tilde{b}} \nabla^{p} \right. \nonumber\\ 
&&\left.+ u_2  s^{ -b-\tilde{b}} 
+ u_3 s^{- 2b-\tilde{b}} \xi + u_4 s^{- 3b-\tilde{b}} \xi^2 -\frac{\Gamma}{2} s^{- b-2\tilde{b}} \tilde{\xi} \right] \xi\,.
\end{eqnarray}
Therefore, we obtain the following relations of the parameters under the scaling transformation:
\begin{eqnarray} \label{eq:rg}
\tau &\to& \tau' = s^{d -b-\tilde{b} }\,\tau \,, \nonumber\\
D &\to& D' = s^{d +z-2-b-\tilde{b} }\,D \,, \nonumber\\
D_p &\to& D'_p = s^{d +z-p-b-\tilde{b} }\,D_p \,, \nonumber\\
u_2 &\to& u'_2 = s^{d +z-b-\tilde{b} }\,u_2\,, \nonumber\\
u_3 &\to& u'_3 = s^{d +z-2b-\tilde{b} }\,u_3 \,, \nonumber\\
u_4 &\to& u'_4 = s^{d +z-3b-\tilde{b} }\,u_4\,, \nonumber\\
\Gamma &\to& \Gamma' = s^{d +z-b-2\tilde{b} } \,\Gamma \,. 
\end{eqnarray}
Note that the transformations of the parameters in Eq.~\eqref{eq:rg}  
correspond to the Wilson renormalization group (RG) procedure~\cite{,hinrichsen,wilson}. 

One can choose $b+\tilde{b} =d$ so that $\tau$ is invariant under the transformation in
Eq.~\eqref{eq:rg}. Moreover, the relations for $D$ and $D_p$ suggest that the dynamic 
exponent $z=2$ for $p>2$ and $z=p$ for $p<2$. If $z=2$ at $p>2$, the long-range term with $D_p$ 
becomes irrelevant, whereas the short-range term with $D$ is relevant, and 
vice versa for $p<2$ with $z=p$. 
Above the upper critical dimension, $d>d_c$, 
the higher-order potential terms and noise term are irrelevant, so
the Gaussian fixed point is stable. Therefore, the relevance of $u_3$, $u_4$, and $\Gamma$ 
determines the upper critical dimension. 
The case of finite $u_3$ is well-known, as follows~\cite{LDP1, LDP2, LDP3, LDP4, LDP5, LDP6}.
If $u_3$ is finite, $u_4$ is automatically irrelevant at $d_c$,
which implies that at $d \lesssim d_c$, $u_3$ and $\Gamma$ are relevant. Thus, one may infer that
at $d=d_c$, $b=\tilde{b}=z$, leading to $b=d_c/2=z$. Because $z=2$ or $z=p$, the upper critical dimensions
of the short-range and long-range QCPs are given by $d_c=4$ and $d_c=2p$, respectively.   
Note that by using $b=\beta/\nu_{\bot}$ with $\beta=1$, one can obtain the MF exponents
obtained in the noiseless equation, Eq.~\eqref{eq:xi3}.

Finally, we discuss the TDP universality with the long-range interaction.
In this case, $u_3=0$; thus, $u_4$ and $\Gamma$ become relevant terms at $d\lesssim d_c$.
Similar to the case of DP, at $d=d_c$, the invariance of $u_4$ and $\Gamma$ in Eq.~\eqref{eq:rg}
yields $b = z/2$ and $\tilde{b}=z$. Because $b+\tilde{b} = d_c$, the upper critical dimension of TDP
is given by $d_c=3z/2$. Therefore, for short-range TDP, it is found that $d_c=3$, as shown 
in previous works~\cite{grassberger_conjecture, lubeck, grassberger3}, and for long-range
TDP with $p<2$, it is found that $d_c=3p/2$, which is similar to the long-range DP case, but
the constant differs from $2$ for the DP class. Again, with $b= \beta/\nu_{\bot}$ and $\beta=1/2$,
we obtain the MF exponents for the TDP universality. 
Because it is well known that the tricritical point does not exist in the one-dimensional DP-type model
in the absence of the long-range interaction~\cite{firstorder1d}, one may ask whether the tricritical point is
sustained when $d=1$ is below the upper critical dimension or $p > 2/3$.
To answer that, numerical studies and RG approaches to long-range TDP are needed.     
    
\section{Discussion and Conclusion}
\label{sec:conclusion}

Now, we discuss how the long-range QCP can emerge from the cold atomic system.
We start with the Hamiltonian of Rydberg atoms under the antiblockade effect~\cite{marcuzzi1}:
\begin{equation}
\hat{H}_{R}=\Omega\sum_l^N \hat{\sigma}_l^x+
\Delta\sum_l^N \hat{n}_l+\sum_{l\neq m}\frac{V_{lm}}{2}\hat{n}_l \hat{n}_m\,,
\label{eq:HR}
\end{equation}
where $\Omega$ is the Rabi oscillation frequency, $\Delta$ denotes the
detuning energy, and $V_{lm}$ is the long-range interaction between excited atoms.
Because $\Delta$ is very large, Rabi oscillation is suppressed, but if we set
$V_{lm}=-\Delta$ for the nearest-neighbor pairs, the excitation can be enhanced
by the interaction. This mechanism leads to coherent and incoherent CPs, where 
the long-range nature of $V_{lm}$ is usually neglected
to realize the absorbing state on long time scales~\cite{marcuzzi1,marcuzzi2,buchhold}. 
However, these approaches are based on the low-density limit; therefore, one spin can interact 
with approximately only one particle. As pointed out in a previous work~\cite{marcuzzi1},
when one spin simultaneously interacts with not only the nearest-neighbor spins, but also long-distance spins, the long-range effect may change the universality of the system.

We investigated the critical behavior of the quantum long-range CP, which is realized by coherently and incoherently driven interacting cold atomic systems. We derived the Heisenberg equations from the total Hamiltonian consisting of the system, the baths, and their interaction. Using the semiclassical approach, we obtained the MF equation for the long-range QCP, where branching and coagulation are realized as L\'evy flight. Then we obtained a phase diagram similar to that for the short-range QCP. Next, we set up the phenomenological Langevin equation and built the Martin--Siggia--Rose--Janssen--de Dominicis action. 
Using scaling theory, we determined the critical exponents in the MF limit. Depending on the model parameters, the DP-type and TDP-type transitions occur. For the DP-type case, the critical exponents were obtained as those of the long-range DP~\cite{LDP1,LDP5}. 
For the TDP-type case at the tricritical point, new critical exponents were obtained, the universality class of which we identify as the long-range TDP class. 
Moreover, we determined the upper critical dimension for the long-range TDP, $d_c=3p/2$, 
which is different from that of the long-range DP class, $d_c=2p$. 
The critical exponents for the ordinary DP and TDP and the long-range DP and TDP classes are compared in Table~\ref{table}.
Recently a similar result that a first and second-order phase transition coexist has been reported in the quantum epidemic model, realizable in a dissipative atomic system with long-range interaction~\cite{epidemic}. 
We expect that our semi-classical approach is also applicable to the epidemic model using a three-state quantum spin system.

In this study, we focused on the long-range nonequilibrium absorbing phase transition 
in the dissipative quantum spin system. We obtained the phase diagram and determined the transition properties within the analytic theoretical framework in the MF limit. However, the transition behavior below the upper critical dimension has not been determined yet. The renormalization group approach to this problem seems to be challenging, yet numerical simulation studies remain as the next problem.

\begin{acknowledgments}
This research was supported by the NRF, Grant No.~NRF-2017R1D1A1B03030872 (JU) and
No.~NRF-2014R1A3A2069005 (BK).
\end{acknowledgments}

\end{document}